# Shape-preserving diffusion of a high-order mode

Dimity Yankelev,[1,*] Ofer Firstenberg,[1,2] Moshe Shuker,[1] and Nir Davidson[3]

[1]Department of Physics, Technion-Israel Institute of Technology, Haifa 32000, Israel
[2]Department of Physics, Harvard University, Cambridge, MA 02138, USA
[3]Department of Physics of Complex Systems, Weizmann Institute of Science, Rehovot 76100, Israel
*Corresponding author: dimitryy@tx.technion.ac.il

The close relation between the processes of paraxial diffraction and coherent diffusion is reflected in the similarity between their shape-preserving solutions, notably the Gaussian modes. Differences between these solutions enter only for high-order modes. Here we experimentally study the behavior of shape-preserving high-order modes of coherent diffusion, known as 'elegant' modes, and contrast them with the non-shape-preserving evolution of the corresponding 'standard' modes of optical diffraction. Diffusion of the light field is obtained by mapping it onto the atomic coherence field of a diffusing vapor in a storage-of-light setup. The growth of the elegant mode fits well the theoretical expectations.

The so-called 'standard' Gaussian modes, such as standard Hermite-Gauss (sHG) and standard Laguerre-Gauss (sLG) modes, form the well-known orthonormal sets of solutions for free-space paraxial propagation and for optical resonators [1]. Analytically, these modes are comprise a real polynomial term, such as the Hermite polynomial for sHG, multiplied by a complex Gaussian,

$$E_{n,m}^{sHG}(x,y,z;w_0) \propto H_n\left(\frac{\sqrt{2}x}{w(z)}\right) H_m\left(\frac{\sqrt{2}y}{w(z)}\right) e^{-(\tilde{x}^2+\tilde{y}^2)}. \quad (1)$$

Here, $H_n$ is the Hermite polynomial of order n and $\tilde{x} = x[ik/2/q(z;w_0)]^{1/2}$ is the complex scaled coordinate, with $q(z;w_0) = z + iz_R$ the complex radius, $z_R = kw_0^2/2$ the Rayleigh length, $k$ the wave number, $w(z) = w_0\sqrt{1+(z/z_R)^2}$ the waist radius, and $w_0$ the waist radius at the waist plane $z=0$. This mathematical form bears a certain asymmetry: the scaling of the spatial coordinates is complex for the Gaussian term but real for the polynomial. In [2], an alternative set of solutions for paraxial diffraction was proposed, having a functional structure similar to the standard HG modes, but with a mutual complex argument for all terms,

$$E_{n,m}^{eHG}(x,y,z;w_0) \propto H_n(\tilde{x})H_m(\tilde{y})e^{-(\tilde{x}^2+\tilde{y}^2)}. \quad (2)$$

These modes were labeled 'elegant' HG (eHG) modes. As can be seen by comparing eq. 1 and eq. 2, the standard and elegant forms are equivalent for the low-order modes ($n=0,1$; $m=0,1$), which are therefore called 'common' modes. Nevertheless, the evolution of the elegant modes in free space is generally not shape-preserving, due to the imaginary part of the polynomial affecting the propagation.

Subsequently, the intensity and phase behavior of the elegant modes were investigated [1], analogues elegant LG modes were derived [3,4], and the near- and far-field evolution was studied [5]. While the elegant modes are inefficient for describing paraxial propagation, they gather interest for various reasons, notably for describing propagation in complex parabolic media [1,6], because of their similarity to high-order Bessel-Gauss beams [7], and in relation to multipole fields from complex point sources [8].

Most recently, the elegant modes were studied within the context of diffusion [9]. Using a storage-of-light technique, the complex electric field of a *probe* light pulse was spatially mapped onto the ground-state coherence (dipoles) of diffusing atoms in a vapor cell with a buffer gas. During storage, the field of atomic coherence $\psi$ undergoes coherent diffusion $\dot{\psi} = D\nabla^2\psi - \gamma\psi$, where $D$ is the diffusion coefficient of the atoms and $\gamma$ an atomic damping rate [10,11]. The retrieved light thus constitutes the input field after effectively experiencing diffusion. For example, the evolution in the waist plane of the elegant HG modes under diffusion was predicted to be [9]

$$E^r(x,y;\tau) = e^{-\gamma\tau}s(\tau)^{-(n+m+1)}E_{n,m}^{eHG}(x,y,w_0s(\tau)), \quad (3)$$

where $E^r$ is the retrieved pulse after storage of duration $\tau$. Notably, eq. 3 describes a shape-preserving evolution of the mode $E_{n,m}^{eHG}$, where the waist radius increases with the scaling factor $s(\tau) = (1+4D\tau/w_0^2)^{1/2}$. The first and second terms account for exponential and geometrical decay, the latter increasing with the mode's order [By definition, $E_{n,m}^{eHG}$ is normalized such that the power $\int dxdy |E_{n,m}^{eHG}|^2$ is constant and does not change with $w_0s(\tau)$.]. However, the shape-preserving evolution under diffusion was tested in [9] only for the common modes.

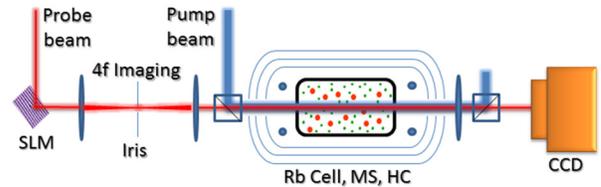

Fig. 1. (Color online.) Scheme of the experimental setup. A probe pulse is reflected from a spatial light-modulator (SLM), which imprints the desired phase pattern. A 4f telescope images the SLM plane to the middle of the Rb cell, with an iris filtering high spatial frequencies. Polarizing beam splitters are used to combine the pump with the probe before the cell and to separate them afterwards. The cell is surrounded by 3 layers of magnetic shielding and by a set of Helmholtz coils for controlling the magnetic field and setting the quantization axis. The retrieved probe pulse after storage is imaged with a single lens to a CCD.

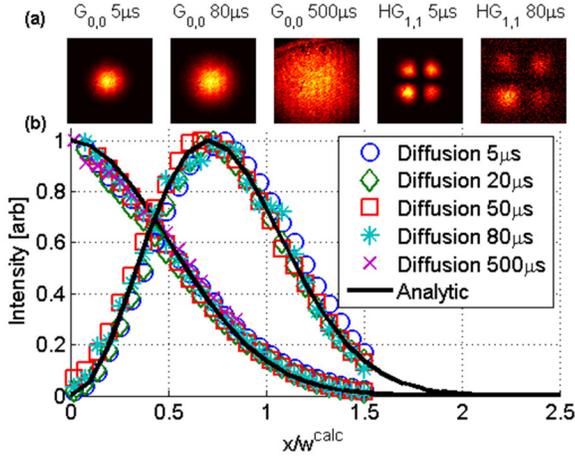

Fig. 2. (Color online.) (a) Images of the Gaussian and $HG_{1,1}$ modes after diffusion, for different storage durations. The intensity maps are normalized. (b) Normalized intensity profiles of the Gaussian and $HG_{1,1}$ modes after diffusion. Solid curves are the analytic expressions for Gaussian and $HG_{1,1}$ modes.

Here we present an experimental and theoretical investigation of the behavior of high-order purely elegant or purely standard Gaussian modes and verify that the standard modes are only shape preserving under diffraction, whereas the elegant ones are only shape-preserving under diffusion.

Our light storage experiment is performed in an electromagnetically-induced transparency setup, depicted in Fig. 1. An external-cavity diode laser in the Littrow configuration [12] is locked at 795 nm to the ($F = 2 \rightarrow F' = 1$) D1 transition of $^{87}Rb$ using a polarization locking scheme [13]. The light is amplified with a tapered amplifier and split into two beams: a 'probe' and a 'pump'. Acousto-optic modulators are used to shape the probe into ~1$\mu s$ long pulse and to turn the pump on and off at desired times, as well as setting the frequency detuning to be equal to the Zeeman splitting between $m = 0$ and $m = 2$ states of the $F = 2$ manifold of the ground state (~50 kHz). After a desired pattern is imprinted on the weak (~10 µW) probe pulse, it is recombined with the strong (~10 mW) pump and sent to the vapor cell. The small probe beam ($w_0 \approx 0.5$ mm) is coaxial with the pump beam ($w_0 \approx 2$ mm), such that the pump intensity for the EIT is effectively uniform. A quarter-wave plate transforms the orthogonal linear polarizations of the probe and the pump into circular polarizations before the cell. The cell contains isotopically pure $^{87}Rb$, with vapor temperature at 50º C and Ne buffer gas pressure of 10 Torr. While the probe slowly propagates through the medium, the pump is turned off, and the resonant probe is coherently absorbed. Turning the pump back on after certain duration coherently excites the probe pulse, which, similarly to holographic retrieval, continues to propagate in its original direction. Finally, the probe is separated from the pump and imaged to a CCD (Sensicam).

To produce the elegant mode, we use a spatial light-modulator (SLM; Holoeye HEO 1080p reflective liquid-crystal) and a 4f imaging system with an iris at the focal plane. A Gaussian probe beam is reflected from the SLM, which displays a computer-generated phase pattern. The basic phase mask consists of π phase flips along straight lines, corresponding to the zero-crossings of the desired HG mode. In addition, we superimpose a blazed-grating pattern to separate the first-order diffraction from the zero order, and a weak lens-like pattern to compensate for phase curvature and to brings the waist plane to the middle of the cell. An iris at the focal plane of the 4f setup filters the high spatial frequencies and blocks the unmodified light. The creation of a given mode consists of iteratively tweaking the distance between the lines of the π phase flips and the aperture of the iris, until the desired intensity image at the cell is achieved.

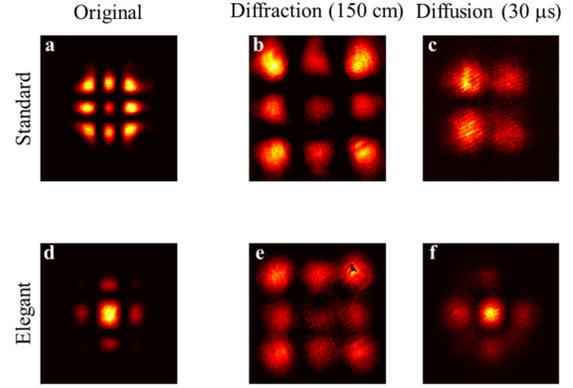

Fig. 3. (Color online.) Comparison between standard and elegant $HG_{2,2}$ modes (a and d, respectively), after diffraction for 150 cm (b and e) or after diffusion for 30 µs (c and f). It can be seen that the standard mode is shape preserving only under diffraction, while the elegant is shape preserving only under diffusion.

Typical images of the diffusion of common modes are shown in Fig. 2(a) for a Gaussian mode (labeled $G_{0,0}$) and for $HG_{1,1}$. For both modes, the diffusing atoms act to stretch the pattern. However for the $HG_{1,1}$, due to the alternating phases between adjacent lobes, atoms that diffuse from the (bright) lobes to the central (dark) regions destructively interfere and do not contribute to the restored light [14]. Fig. 2(b) presents the scaled and normalized cross sections. It can be seen for both modes that the evolution is shape preserving. The diffusion coefficient obtained from fitting these results to the theoretical expectation is $D = 12$ cm$^2$/s. This value is consistent with previous results [9] but is here affirmed for significantly longer storage durations, owing to improved uniformity of the magnetic field.

Next, the diffusion evolution of the elegant and standard $HG_{2,2}$ modes, which are not 'common', is studied in comparison to their diffraction evolution. A qualitative comparison is presented in Fig. 3. The standard $HG_{2,2}$ mode expands in a shape-preserving manner while propagating in free space and experiencing diffraction, but, as expected, its shape is significantly changed after storage and diffusion. Conversely, the elegant $HG_{2,2}$ mode maintains its shape following diffusion, up to spatial scaling, but is significantly changed during free space diffraction.

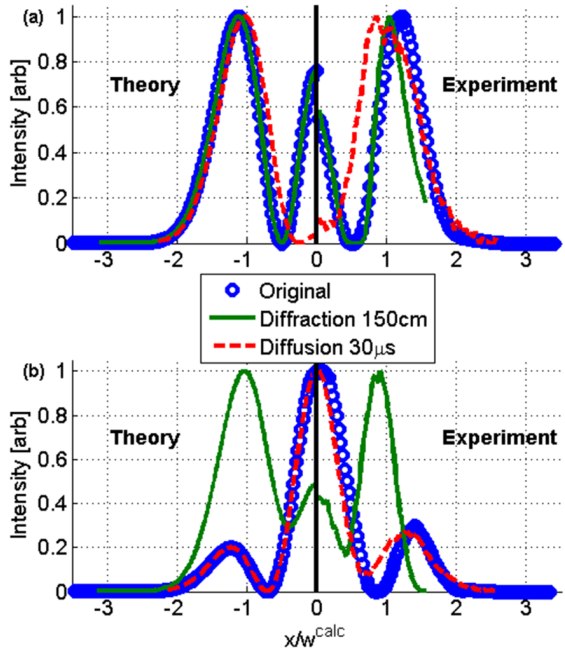

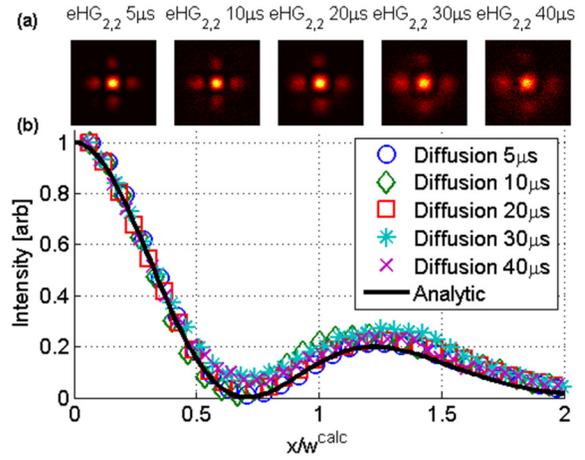

Fig. 4. (Color online.) Cross sections (of the images from Fig. 3) around the maximum intensity of (a) standard and (b) elegant HG$_{2,2}$ modes (blue circles), after diffraction for 150 cm (green solid line) and diffusion for 30 µs (red dashed line). On the left are the theoretical profiles, and on the right the experimental ones. The cross sections are normalized according to the fitted or calculated waist, for the original and the diffracted/diffused results.

The cross sections of the beams are given in Fig. 4, including comparisons to the theoretically calculated curves [Fig. 4(a) for sHG$_{2,2}$ and Fig. 4(b) for eHG$_{2,2}$]. All cross sections are scaled according to the expected scalings: $s(\tau)$ for diffusion (with $\tau = 30$ µs and the previously obtained value of $D$) and the Rayleigh expression $[1 + (2z/kw_0^2)^2]^{1/2}$ (with $z = 50$ cm) for diffraction. The theoretical calculations, which agree very well with the experiment, are done by numerically taking the Fourier transform and calculating the effect of diffusion or diffraction in the Fourier space. The small deviations of the experimental results from theory are attributed to slight initial deviations from the intended mode. In particular, the elegant modes are very sensitive to a curvature of the phase plane, because, in contrast to the standard modes, it does not simply correspond to shifting the focal plane.

The detailed evolution of eHG$_{2,2}$ under diffusion for different storage durations is shown in Fig. 5. The cross sections in Fig. 5(b) are taken around the center and normalized to maximal unit intensity. The spatial coordinate is scaled according to $s(t)$. It is evident that the evolution is shape preserving and that the expansion rate is in good agreement with the theory and with the results of the common modes.

To summarize, shape-preserving solutions of optical diffraction differ from those of diffusion only at high orders. We have verified this experimentally by monitoring the evolution of purely-elegant and purely-standard modes under both processes. The observed large sensitivity of the elegant modes to an initial phase curvature – corresponding to an off-focal imaging for standard mode – is essentially due to the diffusion and diffraction acting simultaneously. Shape-preserving solutions of orders $n \geq 2$ for such combined dynamics are yet to be studied.

Fig. 5. (Color online.) (a) Images of the elegant HG$_{2,2}$ mode after diffusion, for different storage durations. The intensity maps are normalized. (b) Normalized intensity cross-section after diffusion of the elegant HG$_{2,2}$ mode around the center, for different storage durations. Solid curve is the analytic expression for elegant HG$_{2,2}$ mode (Eq. 3).

We gratefully acknowledge discussions with A. Ron, and technical assistance from P. London, O. Peleg, I. Ben-Aroya, and Y. Erlich.